\documentclass[aps,prb,reprint,showpacs,superscriptaddress,floatfix]{revtex4-2}
\usepackage[dvipdfmx]{graphicx}
\usepackage{amsmath}
\usepackage{float}
\usepackage{subfigure}
\usepackage{tikz}
\usepackage[english]{babel}
\usepackage{amssymb}
\usepackage{mathrsfs}
\usepackage{physics}
\usepackage[utf8]{inputenc}
\usepackage{booktabs}
\usepackage{hyperref}
\usepackage[normalem]{ulem}
\usepackage{xcolor}
\definecolor{Black}{named}{black}

\newcommand{\non}{\nonumber}

\def\Hex{H_{\mathrm{ex}}}

\begin{document}

\title{Quantum oscillations of valley current driven by microwave irradiation in transition-metal dichalcogenide/ferromagnet hybrids}
%\title{Microwave-driven quantum oscillations of valley current in transition-metal dichalcogenide/ferromagnet hybrids}
%\title{Quantum oscillation of valley current driven by microwave irradiation in monolayer transition-metal dichalcogenide/ferromagnetic insulator hybrids}%a tentative title
% Quantum oscillation of valley current driven by microwave irradiation in TMDC-FI hybrids 

\author{Xin Hu }
\affiliation{%
Kavli Institute for Theoretical Sciences, University of Chinese Academy of Sciences, Beijing, 100190, China.
}%

\author{Yuya Ominato}
\affiliation{%
Waseda Institute for Advanced Study, Waseda University, Shinjuku-ku, Tokyo 169-0051, Japan.
}%

\author{Mamoru Matsuo }
\email{mamoru@ucas.ac.cn}
\affiliation{%
Kavli Institute for Theoretical Sciences, University of Chinese Academy of Sciences, Beijing, 100190, China.
}%
\affiliation{%
CAS Center for Excellence in Topological Quantum Computation, University of Chinese Academy of Sciences, Beijing 100190, China
}%
\affiliation{%
Advanced Science Research Center, Japan Atomic Energy Agency, Tokai, 319-1195, Japan
}%
\affiliation{%
RIKEN Center for Emergent Matter Science (CEMS), Wako, Saitama 351-0198, Japan
}%

\date{\today}
\begin{abstract}
We theoretically study spin and valley transport in a transition-metal dichalcogenides (TMDC)/ferromagnet heterostructure under a perpendicular magnetic field. We find that microwave-driven spin pumping induces a valley-selective spin excitation, a direct consequence of the valley-asymmetric Landau levels in the TMDC conduction band. This process generates a pure valley current, which, as our central finding, exhibits pronounced quantum oscillations as a function of chemical potential. These oscillations provide a definitive experimental signature of the quantized valley states and establish another pathway to interface spintronics and valleytronics.
\end{abstract}

\maketitle

{\it Introduction.---}Monolayer transition-metal dichalcogenides (TMDCs) have become a central platform for exploring emergent quantum phenomena due to their unique electronic band structure, which intrinsically couples spin and valley degrees of freedom \cite{PhysRevLett.99.236809,PhysRevB.77.235406,RevModPhys.82.1959,PhysRevB.81.195431,PhysRevB.84.125427}. The conduction and valence band edges are located at the distinct $K$ and $K'$ points of the Brillouin zone \cite{PhysRevLett.105.136805, li2007electronic, splendiani2010emerging, PhysRevB.79.115409, zhuGiantSpinorbitinducedSpin2011}, establishing the valley as a robust quantum number. Furthermore, strong spin-orbit interaction in conjunction with broken inversion symmetry gives rise to spin-valley coupling (SVC), locking the spin orientation to a specific valley \cite{ zhuGiantSpinorbitinducedSpin2011,xiaoCoupledSpinValley2012,wang2015strong,zihlmann2018large, wakamura2018strong}. This feature positions TMDCs as a highly promising material for future spintronic and valleytronic devices, where spin and valley currents act as carriers of information and energy \cite{RevModPhys.76.323, PhysRevLett.97.186404, ye2016electrical}.

A variety of methods have been established to access and control the valley degree of freedom in TMDCs. Prominent examples include optical generation via valley-selective circular dichroism~\cite{caoValleyselectiveCircularDichroism2012, mak2012control, PhysRevB.86.081301} and electrical generation through the valley Hall effect or spin-polarized charge injection~\cite{sanchezValleyPolarizationSpin2016, ye2016electrical}. 
These approaches, however, fundamentally rely on charge-based excitations using light or electric fields. While the concept of generating a valley current via pure spin injection has been proposed, theoretical exploration has been limited to systems without an external magnetic field and has primarily focused on the valence band \cite{ominato_valley-dependent_2020}. The application of a perpendicular magnetic field fundamentally alters the electronic landscape by quantizing the energy bands into discrete, spin- and valley-polarized Landau levels \cite{caiMagneticControlValley2013, liUnconventionalQuantumHall2013, wangValleySpinpolarizedLandau2017}. 
The behavior of these quantized levels under pure spin excitation, along with the possibility of novel transport effects, has yet to be investigated and represents a compelling open question.

\begin{figure}[htbp]
    \centering
    \includegraphics[width=0.49\textwidth]{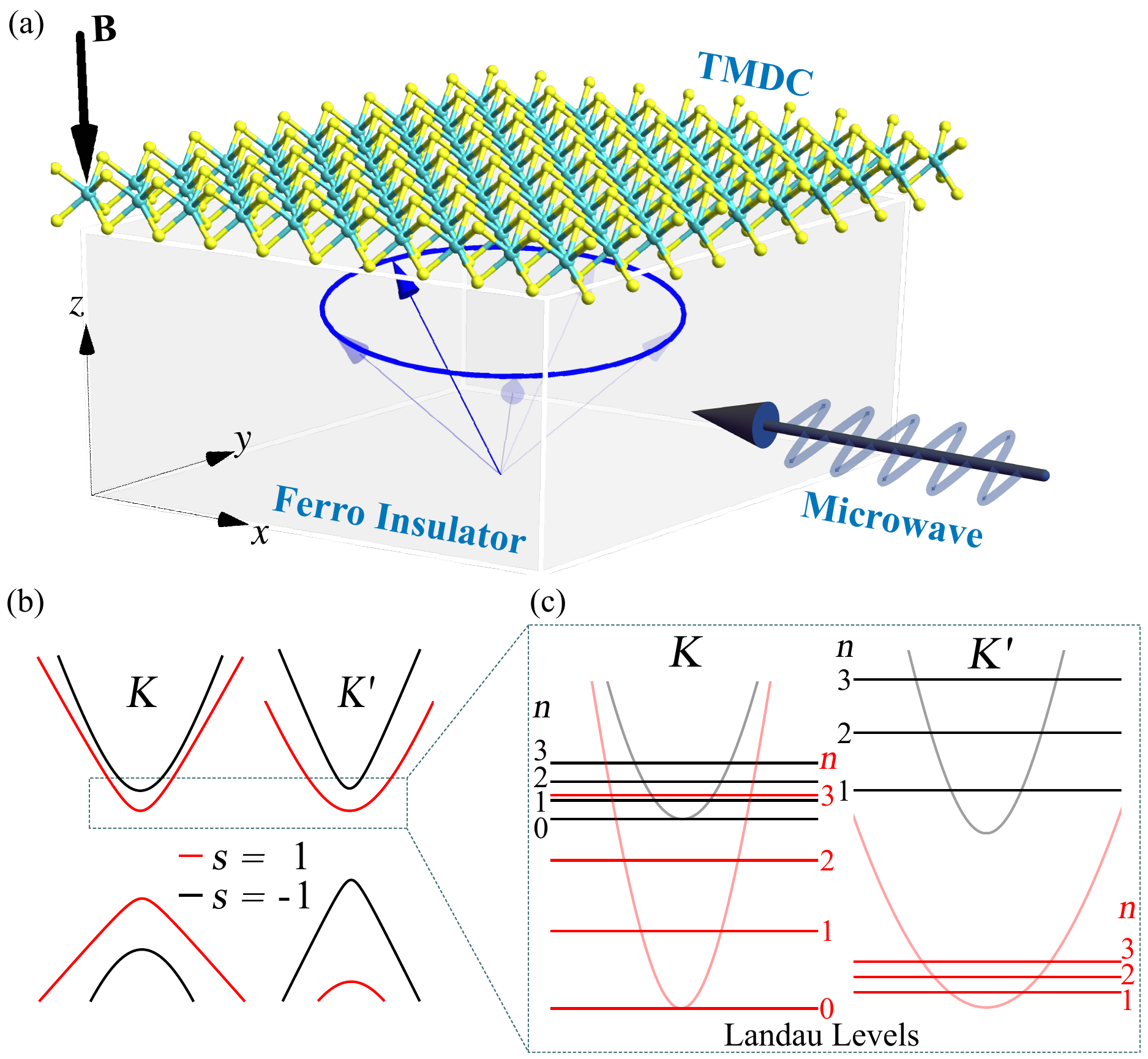}
    \caption{     
    (a) Schematic of the proposed system. A monolayer of transition-metal dichalcogenide (TMDC) is placed on a ferromagnetic insulator (FI). Microwave irradiation induces ferromagnetic resonance in the FI, generating a pure spin current ($I_s$) at the interface via spin pumping. A perpendicular magnetic field ($B$) is applied to the entire heterostructure. (b) Electronic band structure of the TMDC in the absence of a magnetic field, showing the characteristic spin-valley locking at the $K$ and $K'$ points. (c) In the presence of a magnetic field, the energy bands quantize into discrete Landau levels. In the conduction band, there exists a valley-selective energy splitting between the lowest Landau levels of the $K$ and $K'$ valleys as the key mechanism enabling the generation of a pure valley current from the pumped spin current.
    }
    \label{system}
\end{figure}

In this Letter, we theoretically investigate spin and valley transport in a TMDC monolayer subjected to a perpendicular magnetic field, driven by a pure spin excitation. As depicted in Fig.~\ref{system}(a), we consider a heterostructure where a TMDC layer is placed on a ferromagnetic insulator (FI). Microwave-induced ferromagnetic resonance (FMR) in the FI generates a pure spin current at the interface via spin pumping. We demonstrate that the interplay between SVC and the magnetic-field-induced lifting of valley degeneracy enables a valley-selective spin excitation in the conduction band of the TMDC. This mechanism directly generates not only a spin current but also a pure valley current. Most strikingly, we find that the Landau quantization of the electronic states leads to pronounced quantum oscillations in the generated valley current as a function of the chemical potential. Our work establishes an alternative pathway for the all-spin-based generation and control of valley phenomena and predicts a quantum oscillatory effect that directly bridges spintronics with valley physics.

{\it System Hamiltonian.---}\label{system Hamiltonian}
We model the heterostructure depicted in Fig.~\ref{system}, which consists of a TMDC monolayer coupled to an underlying FI. The system is subjected to a perpendicular magnetic field $B$ and microwave irradiation. The total Hamiltonian is expressed as the sum of three components:
\begin{align}
    H = H_{\mathrm{TM}} + H_{\mathrm{FI}} + H_{\mathrm{ex}}.
\end{align}
Here, $H_{\mathrm{TM}}$ describes the electrons in the TMDC, $H_{\mathrm{FI}}$ represents the magnons in the FI driven by the microwave field, and $H_{\mathrm{ex}}$ accounts for the exchange interaction at the TMDC/FI interface.

The electronic properties of the TMDC near the $K$ ($\tau=+$) and $K'$ ($\tau=-$) valleys are captured by the effective Hamiltonian~\cite{xiaoCoupledSpinValley2012}:
\begin{align}\label{eq_TMDC_eff}
    H_{\mathrm{eff}} = v\left( \tau \pi_{x} \sigma^{x} + \pi_{y} \sigma^{y}\right) + \frac{\Delta}{2} \sigma^{z} - \tau s\lambda \frac{\sigma^{z} - 1}{2},
\end{align}
where $\vb*{\pi} = \vb*{p} + e\vb*{A}$ is the kinetic momentum with the vector potential $\vb*{A}$ in the Landau gauge, $v$ is the velocity, $\sigma^{x,y,z}$ are the Pauli matrices acting on the sublattice space, $\Delta$ is the energy gap, and $\lambda$ is the spin splitting at the valence-band top caused by spin-orbit coupling. The index $s=\pm$ denotes the electron spin. These parameters are obtained by fitting to first-principles calculations~\cite{xiaoCoupledSpinValley2012,zhuGiantSpinorbitinducedSpin2011,cheiwchanchamnangijQuasiparticleBandStructure2012, echeverrySplittingBrightDark2016, kormanyos2015k, kormanyosMonolayerMoS22013a, liuThreebandTightbindingModel2013}.

The Hamiltonian for the FI, within the spin-wave approximation, describes magnons under microwave pumping:
\begin{align}
    H_{\mathrm{FI}}(t)=\sum_{\vb*{k}} \hbar \omega_{\vb*{k}} b_{\vb*{k}}^{\dagger} b_{\vb*{k}}-h_{\mathrm{ac}}^{+}(t) b_{{\vb*{k}}=\mathbf{0}}^{\dagger}-h_{\mathrm{ac}}^{-}(t) b_{{\vb*{k}}=\mathbf{0}},
\end{align}
where $b_{\vb*{k}}^{\dagger}$ ($b_{\vb*{k}}$) is the creation (annihilation) operator for a magnon with wave vector $\vb*{k}$ and energy $\hbar\omega_{\vb*{k}}$ given by the Holstein-Primakoff transformation \cite{PhysRev.58.1098} and employing the spin-wave approximation: $S_{\vb*{k}}^+\simeq\sqrt{2S}b_{\vb*{k}}$ where $S_{\vb*{k}}^{\pm}$ are the Fourier components of $\vb*{S}^{\pm}_j$ as the localized spin in the FI at site $j$.
The second and third terms describe the coupling to a uniform microwave field of frequency $\Omega$, where $h_{\mathrm{ac}}^{\pm}(t) = (\hbar \gamma h_{\mathrm{ac}}/2)\sqrt{2SN} e^{\mp i\Omega t}$, where $h_{\mathrm{ac}}$ is the amplitude, $\gamma$ is the gyromagnetic ratio, $S$ is the spin of the FI, and $N$ is the number of spins in the FI layer.

The exchange interaction at the interface, $\Hex$, is decomposed into a static Zeeman-like term $H_Z$ \cite{PhysRevB.92.121403, zhang2016large, liang2017magnetic, PhysRevB.97.041405, PhysRevB.96.085411, PhysRevLett.122.086401} and a dynamic tunneling term $H_T$~\cite{kato_microscopic_2019,PhysRevLett.120.037201, ohnumaTheorySpinPeltier2017, ohnuma_enhanced_2014}:
\begin{align}
    &\Hex
    =-\int d\vb*{r}\sum_nJ(\vb*{r},\vb*{r}_j)\vb*{s}(\vb*{r})\cdot\vb*{S}_j
    =H_Z+H_T, \\
    &H_Z=-\int d\vb*{r}\sum_jJ(\vb*{r},\vb*{r}_j)s^z(\vb*{r})S^z_j \simeq -J_0Ss^z_{\mathrm{tot}}, \\
    &H_T=-\frac{1}{2}\int d\vb*{r}\sum_j
    J(\vb*{r},\vb*{r}_j)
    \left[s^+(\vb*{r})S^-_j+s^-(\vb*{r})S^+_j\right],
\end{align}
where $s^z_{\mathrm{tot}}$ is the total z-component of the electron spin in the TMDC, and $s^\pm(\vb*{r})$ are the ladder operators. The term $H_Z$ arises from the static part of the exchange coupling, inducing a proximity-exchange field, while $H_T$ describes the spin-flip processes between TMDC electrons and FI magnons, which drives the spin pumping.

We treat $H_T$ as a perturbation to the unperturbed Hamiltonian $H_0 = H_{\mathrm{TM}} + H_{\mathrm{FI}} + H_Z$. The effect of $H_Z$ is to introduce a spin-dependent energy shift $-sJ_0S$ to the electronic states. Solving for the eigenvalues of $H_{\mathrm{eff}} + H_Z$ in a magnetic field yields the Landau level energies:
\begin{align}
\varepsilon_{n,\tau,s}=\varepsilon_0\operatorname{sgn}_{\tau}(n) \sqrt{|n|+(\Delta')^2}+\frac{\tau s \lambda}{2}-sJ_0S_0,
\end{align}
where $\varepsilon_0 = \sqrt{2}\hbar v/l_B$ with the magnetic length $l_B=\sqrt{\hbar/eB}$, and $\Delta' = (\Delta/2 - \tau s \lambda/2)/\varepsilon_0$. The function $\mathrm{sgn}_\tau(n)$ is defined as $+1$ for $n>0$, $-1$ for $n<0$, and $\tau$ for $n=0$.

The resulting Landau level structure for the conduction band is illustrated in Fig.~\ref{system}(c). Crucially, the valley degeneracy is lifted. For instance, the lowest energy state in the $K$ valley corresponds to the spin-up $n=0$ Landau level, whereas for the $K'$ valley, it is the spin-up $n=1$ level. This valley-asymmetric energy splitting is the essential ingredient that allows for the generation of a pure valley current via valley-selective spin excitation.

{\it Valley-dependent spin current at the interface.---}
Microwave irradiation induces FMR in the FI, which in turn excites spin-flips in the TMDC via the exchange coupling at the interface. This process, known as spin pumping, generates a pure spin current flowing from the FI into the TMDC~\cite{mizukami2001study,mizukami2001ferromagnetic,mizukami2002effect,tserkovnyak2002enhanced,RevModPhys.76.323,tserkovnyak2005nonlocal,Hellman2017-fm}. The operator for the $z$-component of the spin current is defined as the rate of change of the total electron spin in the TMDC: 
\begin{align}
    I_s^z:=-\frac{\hbar}{2}\dot{s}^z_{\mathrm{tot}}=\frac{i}{2}[s^z_{\mathrm{tot}},H].
\end{align}
We calculate the statistical average of the spin current within the second order perturbation theory. Consequently, the spin current is given by \cite{ohnumaTheorySpinPeltier2017,hu_spin_2024,ominato_valley-dependent_2020,kato_microscopic_2019, Ominato2022-xc, Ominato2022-se,PhysRevLett.120.037201, ohnuma_enhanced_2014}:
\begin{align}\label{spin_current}
    \langle I_s^z\rangle
    =\frac{\hbar J_0^2A}{2}\int \frac{d\omega}{2\pi}
    &\mathrm{Im}\chi^R_{\mathrm{loc}}(\omega)
    \mathrm{Im}[\delta G^<_{\mathrm{loc}}(\omega)],
\end{align}
where $A$ is the interface area and $J_0^2$ represents the configurationally averaged exchange coupling~\cite{hu_spin_2024, Ominato2022-se, Ominato2022-xc}.

The first key component, $\chi^R_{\mathrm{loc}}(\omega)$, is the local dynamic spin susceptibility of the TMDC electrons, defined as:
\begin{align}
    \chi^R_{\mathrm{loc}}(\omega):=\int dte^{i(\omega+i0)t}\frac{i}{\hbar}\theta(t)\langle[s^+(\vb*{0},t),s^-(\vb*{0},0)]\rangle_0 ,
\end{align}
where the average $\langle \cdots \rangle_0$ is taken over the unperturbed Hamiltonian. Its imaginary part, which quantifies the density of available spin-flip excitations, is calculated as:
\begin{align}\label{chi}
    \operatorname{Im} \chi _{\mathrm{loc}}^{R}( \omega )=& {\pi \hbar \omega }\sum _{\tau }\int d\varepsilon  \left( -\frac{\partial f_\mathrm{FD}( \varepsilon )}{\partial \varepsilon }\right) W(\varepsilon,\tau) \non \\ 
   & \times  D_{\tau ,+1}^\mathrm{LL}( \varepsilon ) D_{\tau ,-1}^\mathrm{LL}( \varepsilon ),
\end{align}
where $f_{\mathrm{FD}}(\varepsilon)$ is the Fermi-Dirac distribution. The density of states (DOS) for the Landau levels, $D_{\tau,s}^\mathrm{LL}(\varepsilon)$, is defined using a Gaussian function to account for level broadening $\Gamma$~\cite{doi:10.1021/ed044p432,GaussianLorentzianSumProduct2018}:
\begin{align}
    D_{\tau ,s}^\mathrm{LL}( \varepsilon ) \simeq \frac{1}{2\pi l_{B}^{2}}\sum _{n}\frac{1}{\Gamma\sqrt{\pi} }  \exp[-\frac{(\varepsilon -\varepsilon _{n,\tau ,s})^2}{2\Gamma^2}].
\end{align}
The function $W(\varepsilon,\tau)$ represents the matrix element for spin-flip transitions, which depends on the orbital character of the wave functions:
\begin{align}
    W( \varepsilon ,\tau ) =\frac{1}{2} +\frac{1}{8}\frac{\Delta ^{2} -\lambda ^{2}}{\left( \varepsilon -\frac{\tau \lambda }{2} +J_{0} S_{0}\right)\left( \varepsilon +\frac{\tau \lambda }{2} -J_{0} S_{0}\right)},
\end{align}
which stems from $|\phi_{p}^{\dagger}\phi_q|^2$ as the overlap of the eigenfunction of Eq.~(\ref{eq_TMDC_eff}).

The second component of Eq.~(\ref{spin_current}) is the local magnon response in the FI, $\delta G^<_{\mathrm{loc}}(\omega)$. This term is the non-equilibrium part of the local lesser Green's function, $G^<_{\mathrm{loc}}(\omega) := (1/N)\sum_{\vb*{k}}G^<(\vb*{k},\omega)$, where the magnon propagator is defined as $G^<(\vb*{k},\omega) := \int dt e^{i \omega t}(2 S / i \hbar)\langle b_{\vb*{k}}^{\dagger}(0) b_{\vb*{k}}(t)\rangle_0$.
Following perturbation theory, the response is dominated by the uniform mode ($\vb*{k}=\mathbf{0}$) resonantly excited by the microwave field~\cite{kato_microscopic_2019,ominato_valley-dependent_2020,ohnuma_enhanced_2014}:
\begin{align}
    \operatorname{Im}\left[\delta G^{<}(\vb*{k},\omega)\right] = -\frac{1}{\hbar} g(\omega) \delta(\omega-\Omega) \delta_{\vb*{k}, \mathbf{0}}.
\end{align}
Here, the delta function $\delta(\omega-\Omega)$ indicates that the magnon excitation is resonantly enhanced at the microwave frequency $\Omega$. The absorption is described by the dimensionless function $g(\omega)$,
\begin{align}
    g(\omega) = \frac{2 \pi N\left(S \gamma h_{\mathrm{ac}}\right)^2}{(\omega-\omega_{\vb*{k}})^2 + (\alpha_\mathrm{G} \omega)^2},
\end{align}
where $\alpha_\mathrm{G}$ is the dimensionless Gilbert damping constant of the FI~\cite{cherepanovSagaYIGSpectra1993,jinTemperatureDependenceSpinwave2019,kasuyaRelaxationMechanismsFerromagnetic1961}. The term $\delta_{\vb*{k}, \mathbf{0}}$ signifies that the local magnon propagator is contributed primarily by the uniform precession mode.

{\it Quantum oscillations of valley current---}
The total spin susceptibility in Eq.~(\ref{chi}) is a sum over contributions from the $K$ ($\tau=1$) and $K'$ ($\tau=-1$) valleys. This allows us to define a valley-decomposed spin susceptibility,
\begin{align}
    \operatorname{Im}\chi _{\tau,\mathrm{loc}}^{R}( \omega )=& {\pi \hbar \omega }\int d\varepsilon  \left( -\frac{\partial f_{FD}( \varepsilon )}{\partial \varepsilon }\right) W(\varepsilon,\tau) \non \\
& \times D_{\tau ,+1}^{LL}( \varepsilon ) D_{\tau ,-1}^{LL}( \varepsilon ).
\end{align}
Consequently, the total spin current $\langle I_s^z\rangle$ can be expressed as the sum of two valley-polarized components:
\begin{align} 
    \langle I_s^z\rangle=I_{s,\tau=1} + I_{s,\tau=-1},
\end{align}
where we introdcue the valley-polorized spin current $I_{s,\tau}$ as
\begin{align}
    I_{s,\tau}=\frac{\hbar J_0^2A}{2}\int \frac{d\omega}{2\pi}
    &\mathrm{Im}\chi^R_{\tau,\mathrm{loc}}(\omega)
    \mathrm{Im}[\delta G^<_{\mathrm{loc}}(\omega)].
\end{align}
The generation of such valley-polarized spin currents signifies that the spin excitation by magnons is valley-selective. This enables the generation of a pure valley current, defined as the difference between the two spin current components:
\begin{align}
    I_v^z=I_{s,K} - I_{s,K'}.
\end{align} 

To demonstrate the generation of a finite valley current, we perform numerical simulations for two representative TMDC materials, WSe$_2$ and MoS$_2$, using parameters derived from first-principles calculations~\cite{xiaoCoupledSpinValley2012,zhuGiantSpinorbitinducedSpin2011}. The parameters are summarized in Table~\ref{para}.
\begin{table}[htbp]
\centering
\begin{tabular}{l c c c c}
\toprule 
\toprule 
          & {$a$}   & {$t$} & {$\Delta$}  & {$\lambda$}  \\
\midrule
WSe$_2$   & 0.3310  & 1190  & 1600        & 230  \\
MoS$_2$   & 0.3193  & 1100  & 1660        & 75   \\
\bottomrule
\bottomrule
\end{tabular}
\caption{
Material parameters for WSe$_2$ and MoS$_2$. The velocity is given by $v = \frac{\sqrt{3}}{2}\frac{at}{\hbar}$. Units are nm for lattice constant $a$, and meV for hopping parameter $t$, energy gap $\Delta$, and spin-orbit coupling $\lambda$.}
\label{para}
\end{table}

\begin{figure*}[htbp]
    \centering
    \includegraphics[width=0.99\textwidth]{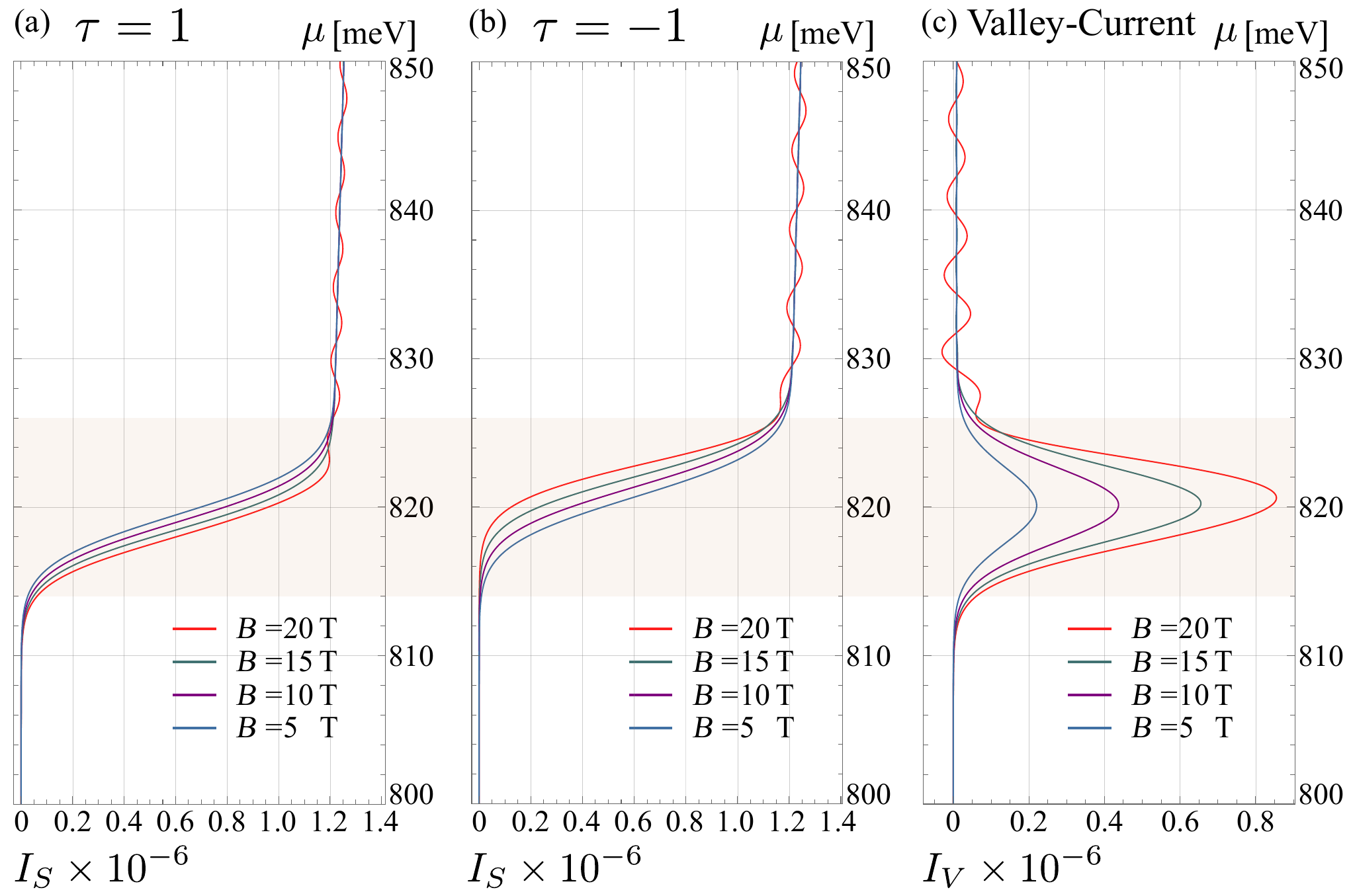}
    \caption{Valley transport in a WSe$_2$/FI heterostructure as a function of chemical potential $\mu$ for magnetic fields $B=$ 5 to 20 T. (a) spin current from the $K$ valley, $I_{s,K}$, (b) spin current from the $K'$ valley, $I_{s,K'}$, and (c) the resulting pure valley current, $I_v$. Simulation parameters: $k_B T=1$ meV, $J_0S_0=20$ meV, $\Gamma=2$ meV.}
    \label{WSe2}
\end{figure*}

\begin{figure*}[htbp]
    \centering
    \includegraphics[width=0.99\textwidth]{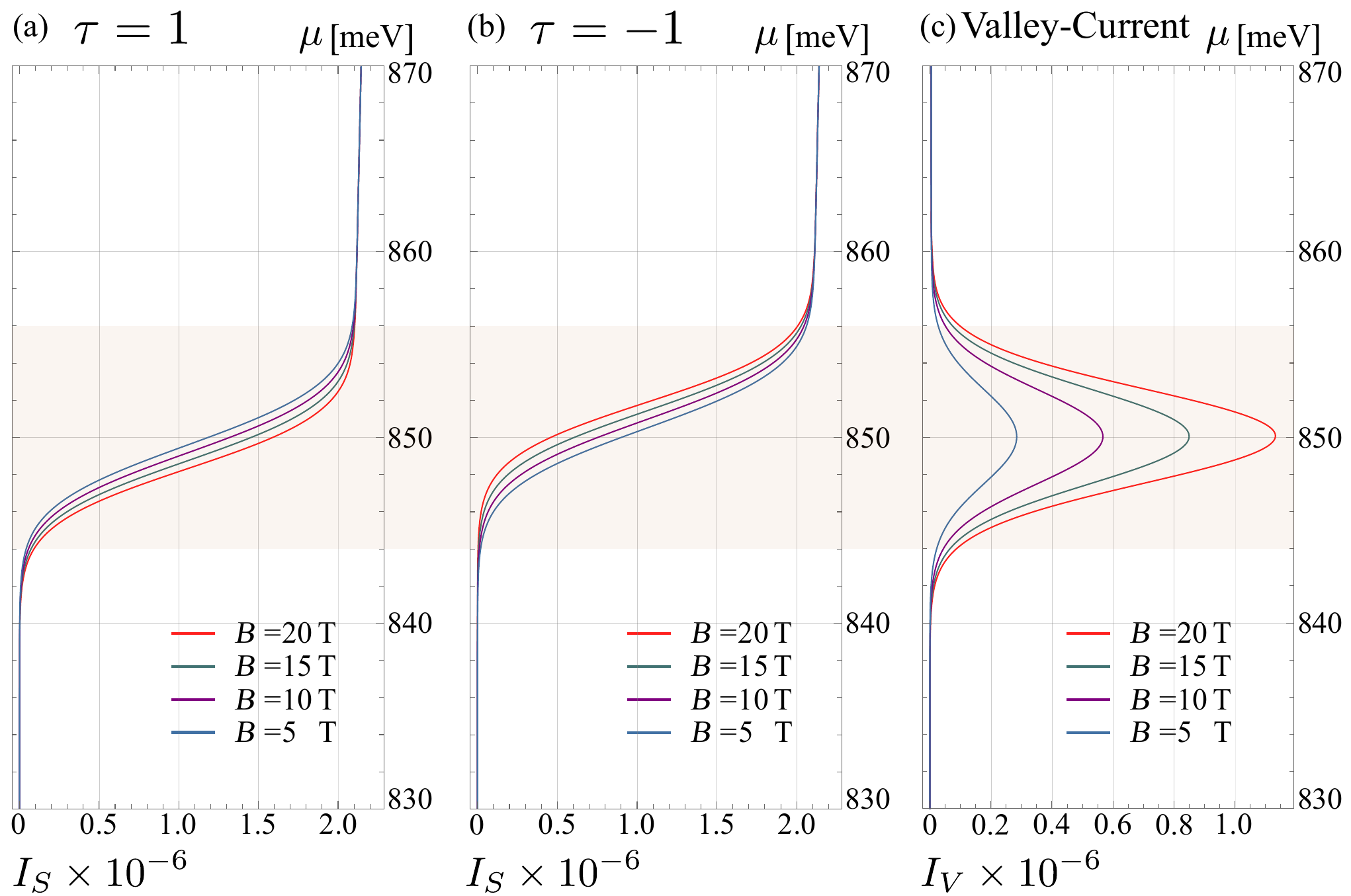}
    \caption{Valley transport in a MoS$_2$/FI heterostructure. Panels and simulation parameters are the same as in Fig.~\ref{WSe2}. The quantum oscillations in the valley current (c) are less pronounced than in WSe$_2$ due to the smaller spin-orbit coupling in MoS$_2$.}
    \label{MoS2}
\end{figure*}

In the following analysis, all currents are normalized by
\begin{align}
    I_0=\frac{\pi \hbar S^2 J_0^2 A }{2}\frac{\Omega ( \gamma h_{\mathrm{ac}} )^2}{( \Omega-\omega _{0})^{2} +\alpha_\mathrm{G}^{2} \Omega^{2}}\frac{1}{(\mathrm{meV})^2(\mathrm{nm})^4}.
\end{align}
Figures~\ref{WSe2} and \ref{MoS2} show the calculated currents for WSe$_2$ and  MoS$_2$, respectively. As seen in panels (a) and (b) of both figures, the spin currents generated from the $K$ and $K'$ valleys exhibit markedly different behaviors as a function of the chemical potential. This asymmetry directly results in a finite pure valley current, $I_v$, as shown in panel (c) of both figures.

The most striking prediction of our theory is the emergence of pronounced quantum oscillations in the valley current, a phenomenon particularly prominent in WSe$_2$ [Fig.~\ref{WSe2}(c)]. This oscillatory behavior is a direct manifestation of the discrete Landau levels traversing the Fermi energy. The oscillations are significantly more pronounced in WSe$_2$ because its larger intrinsic spin-orbit coupling amplifies the valley-asymmetric splitting of the Landau levels. This greater separation between the energy states of the $K$ and $K'$ valleys ultimately produces a more robust and clearly resolved oscillatory signal with increasing magnetic field.

The emergence of a finite valley current, highlighted by the orange shading in Figs.~\ref{WSe2} and \ref{MoS2}, is a direct consequence of the valley-asymmetric Landau level structure. The valley current, defined as the difference between the spin currents from the $K$ and $K'$ valleys, becomes significant because these two components exhibit contrasting behavior with the magnetic field. As shown in Figs.~\ref{WSe2}(a) and \ref{MoS2}(a), the spin current from the $K$ valley, $I_{s,K}$, exhibits enhancement with increasing magnetic field strength. This is because the unique $n=0$ Landau level in the $K$ valley remains field-independent while its degeneracy increases with the field, amplifying the spin-flip response. In contrast, for the $K'$ valley, which lacks the $n=0$ level, the lowest Landau level ($n=1$) shifts to higher energies as the field increases. This shift suppresses the spin-flip transition probability, leading to a decrease in $I_{s,K'}$ as seen in Figs.~\ref{WSe2}(b) and \ref{MoS2}(b).

{\it Discussion.---}
Our proposed mechanism for generating a valley current in the conduction band can be contrasted with related phenomena in the valence band. Although Landau quantization also occurs in the valence band, the energy separation between levels is typically so small that on the same order of magnitude as the level broadening $\Gamma$, making quantum oscillations difficult to resolve. This highlights the distinct advantage of probing the conduction band.

Furthermore, the effect we predict is fundamentally distinct from a similar phenomenon in the valence band of TMDCs~\cite{ominato_valley-dependent_2020}. That work relied on a valley-selective spin excitation enabled by an energy gap induced by the proximity exchange splitting $J_0S_0$, which allows for a valley current even at zero magnetic field. In contrast, our mechanism hinges on the valley-asymmetric structure of Landau levels—specifically, the presence of the $n=0$ level level in only one valley. This requires a finite magnetic field, while the exchange splitting $J_0S_0$ is not essential. The quantum oscillations we predict would survive even in the absence of exchange splitting $J_0S_0$.

We emphasize that our mechanism is not ESR that directly excites conduction electrons in a monolayer TMDC but a TMDC and ferromagnet hybrid driven at ferromagnetic resonance, where the microwave field excites the ferromagnet and interfacial exchange induces spin flips in the TMDC, yielding valley polarized spin tunneling.  
The dc spin current is governed by three elements: (i) the squared interfacial exchange, (ii) the TMDC dynamical spin susceptibility, and (iii) the ferromagnet dynamical spin susceptibility at ferromagnetic resonance. Although ESR in TMDCs has been observed in the GHz range \cite{yang2015spin,stesmans2016esr}, without the ferromagnet, there is no interfacial spin pumping or magnon-driven valley-selective injection. This hybrid coupling of microwave-driven magnon dynamics to valley-resolved electronic states produces a valley-polarized spin current.

Finally, we address the experimental feasibility. Observing these quantum oscillations requires magnetic fields of 5–20 T and a method to inject a pure spin current into the TMDC under such conditions. Magnetic fields of this magnitude are experimentally accessible and have been used in transport measurements on TMDC devices~\cite{ImagingQuantumSpin,kerdiHighMagneticField2020,liUnconventionalQuantumHall2013}. The primary challenge lies in the spin injection. Conventional spin pumping driven by FMR is technically demanding in high-field regimes. Therefore, experimental verification of our predictions will likely necessitate either advancements in high-field FMR spectroscopy or the development of alternative spin injection techniques compatible with strong magnetic fields. 

Regarding modeling assumptions and scope, our analysis employs the simplest two-band Dirac-type description for the TMDC conduction band. More complete treatments indicate that valley splittings can be smaller, especially in WSe2, with quantitative values sensitive to the methodology used to extract valley $g$-factors\cite{kormanyos2015landau}. These refinements and Landau-level broadening constrain the practical window for observation, highlighting material choice, disorder control, and interface engineering as priorities for future work.

{\it Conclusion.---}We have theoretically investigated that spin pumping into a TMDC monolayer under a magnetic field provides a robust pathway for generating a pure valley current. Our central prediction is the emergence of pronounced quantum oscillations in this current, a direct signature of the quantized, valley-asymmetric electronic structure. This phenomenon, particularly prominent in materials with strong spin-orbit coupling such as WSe$_2$, offers a definitive experimental fingerprint. The observation of this effect would provide a direct link between the quantum nature of valleytronics and the versatile field of spintronics, paving the way for novel spin-valleytronic device concepts.

{\it Acknowledgements.--}
This work was supported by the National Natural Science Foundation of China (NSFC) under Grant No. 12374126, by the Priority Program of the Chinese Academy of Sciences under Grant No. XDB28000000, and by JSPS KAKENHI under Grants No.21H01800, No.21H04565, No.23H01839, and No.24H00322 from MEXT, Japan.

\bibliographystyle{modified-apsrev4-2_new.bst}
\bibliography{ref.bib}

\end{document}